\documentclass[journal]{IEEEtran}
\usepackage{amsmath,amsfonts}
\usepackage{algorithmic}
\usepackage{algorithm}
\usepackage{array}
\usepackage[normalem]{ulem}
\usepackage{pifont}
\usepackage{adjustbox}
\usepackage{textcomp}
\usepackage{stfloats}
\usepackage{url}
\usepackage{verbatim}
\usepackage{graphicx}
\usepackage{cite}
\usepackage{acronym}
\usepackage{makecell}
\usepackage{subfigure}

\usepackage[hidelinks]{hyperref}

\usepackage{ifpdf}
\usepackage{url}
\ifpdf
\else
\fi

\usepackage{cite}
\usepackage{balance}
\usepackage{bm,comment,color}

\usepackage{soul}
\usepackage{amsmath}
\usepackage{array}
\usepackage{amsfonts} 
\usepackage{amssymb}
\usepackage{esint} 
\usepackage{units}
\usepackage{multirow}
\usepackage{comment}

\usepackage[table]{xcolor}
\usepackage{hhline}

\usepackage{color}

\usepackage{pgfplots}
\usepackage{tikz}
\usetikzlibrary{calc}
\makeatletter
\newcommand{\gettikzxy}[3]{%
  \tikz@scan@one@point\pgfutil@firstofone#1\relax
  \edef#2{\the\pgf@x}%
  \edef#3{\the\pgf@y}%
}
\usetikzlibrary{spy,backgrounds}

\pgfplotsset{compat=newest}
\usetikzlibrary{plotmarks}
\usetikzlibrary{arrows.meta}
\usepgfplotslibrary{patchplots}
\usepackage{grffile}
\newlength\fheight 
\newlength\fwidth 
\usepgfplotslibrary{fillbetween}

\acrodef{6g}[6G]{the sixth generation}
\acrodef{aoa}[AOA]{angle-of-arrival}
\acrodef{aod}[AOD]{angle-of-departure}
\acrodef{bs}[BS]{base station}
\acrodef{bse}[BSE]{beam squint effect}
\acrodef{elaa}[ELAA]{extremely large antenna array}
\acrodef{ff}[FF]{far-field}
\acrodef{isac}[ISAC]{integrated sensing and communication}
\acrodef{las}[L\&S]{localization and sensing}
\acrodef{los}[LOS]{line-of-sight}
\acrodef{nf}[NF]{near-field}
\acrodef{nlos}[NLOS]{non-line-of-sight}
\acrodef{ofdm}[OFDM]{orthogonal frequency division multiplexing}
\acrodef{ris}[RIS]{reconfigurable intelligent surface}
\acrodef{sns}[SNS]{spatial non-stationarity}
\acrodef{swm}[SWM]{spherical wave model}
\acrodef{siso}[SISO]{single-input single-output}
\acrodef{ue}[UE]{user equipment}
\acrodef{dmimo}[D-MIMO]{distributed MIMO}
\long\def\comment#1{}

\newfont{\bbb}{msbm10 scaled 700}

\newcommand{\hthickline}{\noalign{\hrule height 0.80pt}}

\newfont{\bb}{msbm10 scaled 1100}

\newcommand{\pv}{{\bf p}}

\newcommand{\uv}{{\bf u}}

\newcommand{\Rm}{{\bf R}}

\makeatletter
\def\ps@IEEEtitlepagestyle{%
\def\@oddfoot{\mycopyrightnotice}%
\def\@evenfoot{}%
}
\def\mycopyrightnotice{%
{
} 
\gdef\mycopyrightnotice{}
}

\makeatletter
\let\old@ps@headings\ps@headings
\let\old@ps@IEEEtitlepagestyle\ps@IEEEtitlepagestyle
\def\confheader#1{
\def\@oddhead{\strut\hfill#1\hfill\strut}%
}
\makeatother
\confheader{\scriptsize \shortstack{This work has been submitted to the IEEE for possible publication. Copyright may be transferred without notice, after which this version may no longer be accessible.}
}

\begin{document}

\title{Calibration in RIS-aided Integrated Sensing, Localization and Communication Systems}

\author{
Reza~Ghazalian,
Pinjun~Zheng,
Hui~Chen,
Cuneyd~Ozturk,
Musa~Furkan~Keskin, 
Vincenzo~Sciancalepore,
Sinan~Gezici,
Tareq~Y.~Al-Naffouri,
and~Henk~Wymeersch

\thanks{This paper is partially supported by the SNS JU project 6G-DISAC under the EU's Horizon Europe research and innovation Program under Grant Agreement No 101139130 as well as the Swedish Research Council under project No. 2022-03007, the King Abdullah University of Science and Technology (KAUST) Office of Sponsored Research (OSR) under Award ORA-CRG2021-4695, the European Union under the Italian National Recovery and Resilience Plan
(NRRP) of NextGenerationEU, partnership on “Telecommunications of the Future” (PE00000001 - program “RESTART”), and the EU Horizon project TIMES (Grant no. 101096307).}
}

\maketitle

\begin{abstract}
Reconfigurable intelligent surfaces (RISs) are key enablers for integrated sensing and communication (ISAC) systems in the 6G communication era. With the capability of dynamically shaping the channel, RISs can enhance communication coverage. Additionally, RISs can serve as additional anchors with high angular resolution to improve localization and sensing services in extreme scenarios. However, knowledge of anchors' states such as position, orientation, and hardware impairments are crucial for localization and sensing applications, requiring dedicated calibration, including geometry and hardware calibration. This paper provides an overview of various types of RIS calibration, their impacts, and the challenges they pose in ISAC systems.
\end{abstract}


\section{Introduction}

In 6G mobile communication systems, integration of sensing and communication will enable high-accuracy and high-resolution sensing and communication in a single system, benefiting both functions, with localization also being a part of the sensing functionality. Leveraging high frequencies, wide bandwidth, and large antenna arrays, the network can sense and enable advanced services such as precise localization, posture and gesture recognition, object tracking, and environment reconstruction~\cite{hu2020reconfigurable}. The \acp{ris} technology plays a crucial role in \ac{isac} systems by allowing the shaping and optimization of the wireless channel for both communication and sensing purposes. RISs are typically constructed as a planar surface comprising a large number of controllable elements that can alter the incident electromagnetic (EM) wave's characteristics, such as phase, amplitude, frequency, or even polarization. By properly controlling the configuration of RISs, they can improve the achievable rate and reliability of wireless communication \cite{basar2019wireless}, and enhance sensing and localization  by providing additional geometric constraints. In addition to the previously mentioned key features of RISs, their energy efficiency, lightweight design, ease of deployment, and compatibility with existing wireless infrastructures amplify their significance in next-generation wireless systems, particularly \acs{isac}.

Several types of RISs have been introduced in literature, including passive RIS (PRIS), hybrid RIS (HRIS), active RIS (ARIS), simultaneous transmission and reflection (STAR) RIS, and beyond diagonal RIS (BD-RIS)~\cite{Ahmed2024Active}. Each of these types has its own advantages and disadvantages when applied in localization, sensing and communication systems. These factors may include performance, signal control overhead, and complexity of signal processing algorithms. In addition, each type of RIS has its own exclusive features. For instance, passive RISs have low cost and energy consumption, but they require a complex algorithm for channel estimation. In contrast, HRISs with sensing capability can reduce the complexity of the algorithm; however, they need power splitting ratio optimization to enhance the system's performance. Therefore, each type of RIS can be adopted depending on the scenarios and applications.  However, the implementation of these RISs in existing works requires knowledge of RISs' geometry and hardware states (e.g., position, orientation, RIS phase configuration, and RIS steering vectors). Consequently, calibration in all these aspects is one of the most crucial and challenging tasks in practice\cite{mano1982method,zhao2012localization}, yet it is often neglected in theoretical studies on RIS.

\begin{figure*}[t]
\centering
\centerline{\includegraphics[width=0.7\linewidth]{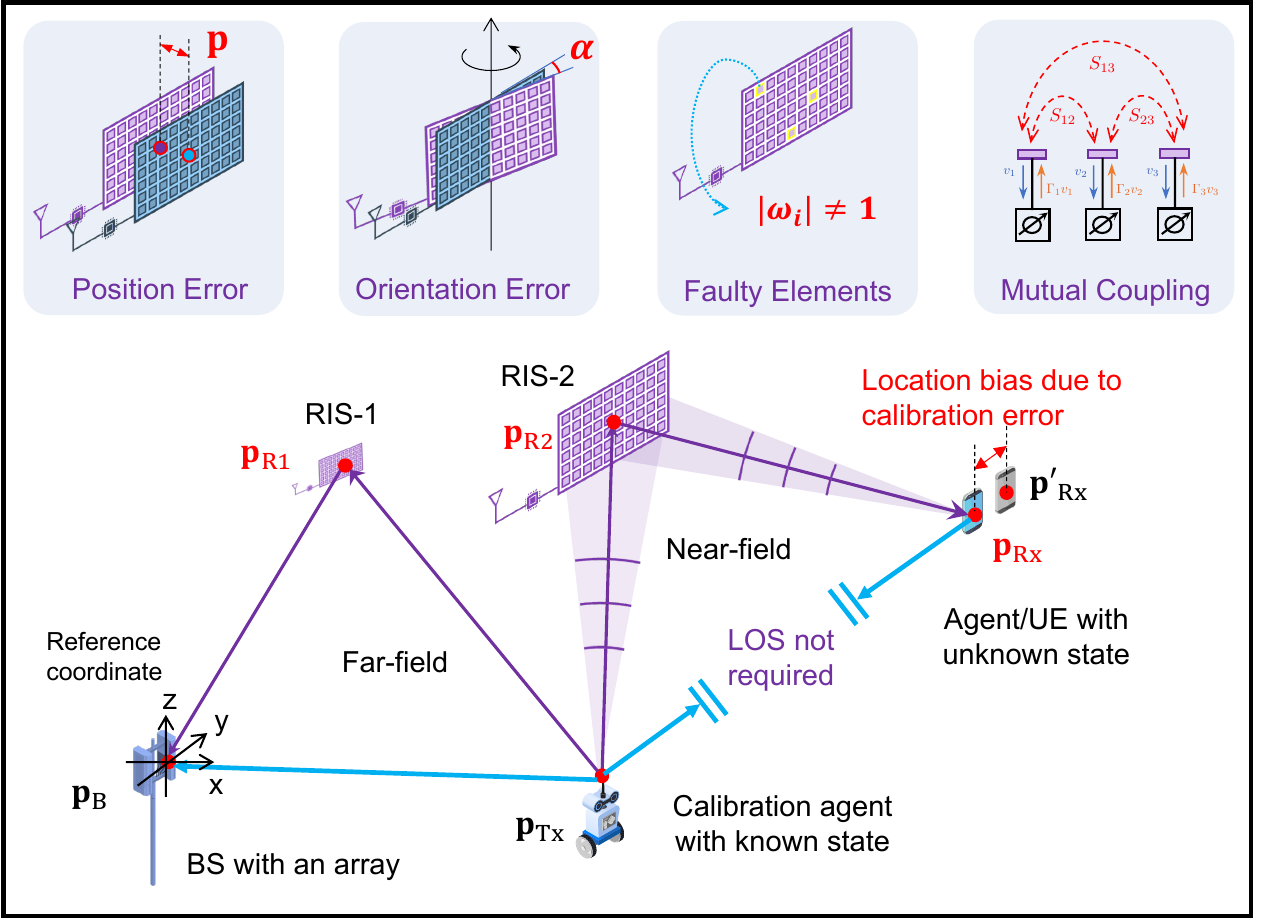}}
\caption{Illustration of calibration in RIS-aided systems. Various effects, such as geometric errors in \ac{ris} position and orientation, faulty elements, and mutual coupling, can lead to model mismatches and potentially degrade localization and communication performance. To fully unleash the potential of RIS-aided systems, agent-based calibration (see RIS-1 where the Tx and Rx states are known) and joint localization and calibration (see RIS-2 where a UE with unknown state is involved for online calibration) can be performed in both far-field and near-field. 
} \vspace{-5mm}
\label{fig-1-illustration}
\end{figure*}

Calibration can be roughly categorized into two types, namely, geometry and hardware. In \textit{geometry calibration}, the position and orientation of the RIS are estimated, which is essential for RIS-aided sensing or localization systems, where RISs serve as reference points. For RIS-aided communication systems, knowledge of the RIS location can facilitate location-based optimization of the RIS phase profile. However, determining the RIS geometric state is non-trivial, especially in indoor applications where users are likely to install RISs without knowing their location or orientation. In \textit{hardware calibration}, we estimate the imperfect states of the RIS hardware (e.g., mutual coupling, pixel failure, and non-linearity of the power amplifier) and compensates for their impacts to prevent performance degradation. Fig.~\ref{fig-1-illustration} presents an overview of these primary types of RIS calibration.

Despite the importance of calibration in RIS-aided \ac{isac} systems, research in this field has  largely overlooked  calibration problems. To address this gap, in this paper, we highlight the negative impact of inaccurate calibration on both localization and communication  performance, examine the effect of the RIS topology on the calibration, and describe the main calibration methodologies. We also present a case study for joint calibration and UE localization. Finally, we discuss future research directions in this area.

\section{Why do we need RIS calibration?}
In this section, we elaborate on two main model mismatches in non-calibrated RIS and then quantify their impact on localization and communication.
\subsection{Model Mismatches}
\subsubsection{Geometric Mismatches}
Obtaining the RIS's position and orientation is inherently challenging in the real world and may introduce additional overhead to the existing signal control for  communication purposes. This overhead stems from the need to control the RIS to effectively measure signals reflected from it or to transfer these measurements, if the RIS has sensing capabilities, to the BS or a local management function. The lack of RIS position or orientation information may be due to several factors, including poor calibration, lack of knowledge about the environment where the RIS is placed, environmental changes, or the mobility of the RIS itself in emerging applications such as RIS-equipped vehicles. These inherent inaccuracies in RIS geometry result in a model where the network operates under the assumption that the RIS is in a specific location with a certain orientation, while in reality, the location and orientation may differ. This, in turn, can lead to  erroneous location references and  beam misalignment. For that reason,  having accurate knowledge of the RIS position and orientation is typically critical for localization and sensing, while it is highly beneficial for communication.

\subsubsection{Hardware Mismatches}

RIS, as a low-cost technology is also susceptible to various hardware-related effects, which typical signal processing methods are unaware of.
\begin{itemize}
    \item  {\emph{Pixel Failure:}} 
In a practical RIS with several hundred unit elements (or pixels), individual elements may be subject to failures. Such pixed failures can result from internal hardware imperfections, such as manufacturing defects, chip interconnection errors and bit flips, and from external physical effects, such as dust, rain and ice~\cite{errorAnalysis_RIS_failure_2020}. While pixel failures due to manufacturing can be detected and calibrated before deploying the RIS, they can still happen unpredictably during normal operation. This situation demands dynamic detection of failures while simultaneously performing localization.

\item {\emph{Mutual Coupling:}} 
In arrays with a half-wavelength or longer inter-element distance, mutual coupling (i.e., the interaction between neighboring elements)  is weak and can safely be ignored. However, in ultra-dense arrays with extremely small inter-element spacing, such as holographic MIMO/RIS, this effect becomes significant and can no longer be neglected~\cite{Zheng2024On}, as it severely affects the RIS radiation pattern. Calibrating mutual coupling in RISs is  more challenging than in conventional antenna arrays, because RIS elements are excited through incident EM waves, which cannot be precisely isolated to a single unit cell for measuring induced voltages at its neighbors. In contrast, conventional antennas are excited through direct feeding ports, allowing the evaluation of coupling parameters by measuring the induced voltages at neighboring antennas when a single antenna is excited.

\item {\emph{Phase-Dependent Amplitude Variations}:}
In RIS reflection, RIS amplitudes are often considered as unity and independent of the phase shift applied at individual elements. However, in real-world RIS prototypes, amplitudes of RIS elements vary as a function of the applied phase shift~\cite{abeywickrama2020intelligent}. To account for this effect in algorithm development and localization performance evaluation, realistic RIS amplitude models based on equivalent circuit models of individual reflecting
elements should be considered. The RIS phase-dependent amplitude variations model in \cite[Eq.~(5)]{abeywickrama2020intelligent} applies to a wide array of semiconductor devices utilized in RIS implementation and is well-supported by experimental findings documented in literature.

\item {\emph{Other Hardware Impairments}:} 
 In addition to the calibration parameters described so far and visualized in Fig.~\ref{fig-1-illustration}, other types of impairments also exist, and some of them only appear in specific RIS architectures. For example, power non-linearity in active RISs, power split coefficient errors, and phase noise introduced in hybrid RISs can occur. Additionally, reflection coefficient errors due to inaccuracies in load impedance considered in circuit design may also affect positioning performance. Consequently, calibration tasks should be well-formulated for different types of RIS systems.
\end{itemize}

\subsubsection{Other Mismatches} 
In addition to geometric and hardware mismatches, there are other mismatches that are not in the focus of this article. These  are related to the channel model, such as beam squint effects or near-field effects. These mismatches do not require calibration methods but instead rely on estimators that account for the correct model. Beam squint, for instance, arises when phase shifters are replaced with phase shifts that vary as a function of frequency in a wideband waveform, resulting in the beam angle changing with frequency. Near-field effects, on the other hand, become significant when the distance between the RIS and the user is relatively short compared to the array size. This requires considering a more complex model and accurate estimation techniques to account for the spherical wavefront.

\subsection{Impact of Model Mismatches on Localization} 
Geometric and hardware model mismatches in the RIS calibration process degrades the performance of both localization and communication systems. To illustrate this point, we provide some relevant examples highlighting the importance of the calibration process. 
Accordingly, to understand the impact of non-calibrated systems, the theory of mismatched estimation provides a useful tool \cite{fortunati2017performance}, wherein the estimator operates under an incorrect or mismatched model. Penalties due to such mismatched operation can be predicted analytically via the misspecified Cramér-Rao bound (MCRB), a generalization of the classical Cramér-Rao bound (CRB).

\subsubsection{Localization Performance under Geometry Mismatch}

\begin{figure}[t]
    \centering
%
%
\definecolor{Magenta}{rgb}{1.00000,0.00000,1.00000}%
\definecolor{ForestGreen}{rgb}{0.1333    0.5451    0.1333}%
\begin{tikzpicture}

\begin{axis}[%
width=2.7in,
height=2.1in,
at={(0in,0in)},
scale only axis,
xmin=-10.5,
xmax=30.5,
xticklabel style = {font=\color{white!15!black},font=\footnotesize},
xlabel style={font=\color{white!15!black},font=\footnotesize},
xlabel={Transmit Power [dBm]},
ymode=log,
ymin=0.01,
ymax=1,
yminorticks=true,
yticklabel style = {font=\color{white!15!black},font=\footnotesize},
ylabel style={font=\color{white!15!black},font=\footnotesize},
ylabel={PEB [m]},
axis background/.style={fill=white},
xmajorgrids,
xminorgrids,
ymajorgrids,
yminorgrids,
grid style=dashed,
legend style={at={(1,1)}, anchor=north east, legend cell align=left, align=left, draw=white!15!black, font=\scriptsize}
]
\addplot [color=Magenta, line width=0.8pt, mark=triangle*, mark options={solid, Magenta}, mark size=3pt]
  table[row sep=crcr]{%
-10	0.868376658524123\\
-5	0.496377374883922\\
0	0.292994210071298\\
5	0.187288159951259\\
10	0.137921592732582\\
15	0.118091026391399\\
20	0.111085739870401\\
25	0.108776624693852\\
30	0.108036147601741\\
};
\addlegendentry{$\text{err}_{\text{pos}}=\unit[0.02]{m},\text{err}_{\text{ori}}=\unit[0.2]{^\circ}$}

\addplot [color=Magenta, dashed, line width=1.1pt, forget plot]
  table[row sep=crcr]{%
-10	0.10769197343643\\
-5	0.10769197343643\\
0	0.10769197343643\\
5	0.10769197343643\\
10	0.10769197343643\\
15	0.10769197343643\\
20	0.10769197343643\\
25	0.10769197343643\\
30	0.10769197343643\\
};
\addplot [color=ForestGreen, line width=0.8pt, mark=square*, mark options={solid, ForestGreen}]
  table[row sep=crcr]{%
-10	0.85467551582234\\
-5	0.484287804995837\\
0	0.278758110325484\\
5	0.167667910301091\\
10	0.1114884242652\\
15	0.0864309733394328\\
20	0.0768246523743201\\
25	0.0735261290505637\\
30	0.0724517981775631\\
};
\addlegendentry{$\text{err}_{\text{pos}}=\unit[0]{m},\text{err}_{\text{ori}}=\unit[0.2]{^\circ}$}

\addplot [color=ForestGreen, dashed, line width=1.1pt, forget plot]
  table[row sep=crcr]{%
-10	0.0719495219497626\\
-5	0.0719495219497626\\
0	0.0719495219497626\\
5	0.0719495219497626\\
10	0.0719495219497626\\
15	0.0719495219497626\\
20	0.0719495219497626\\
25	0.0719495219497626\\
30	0.0719495219497626\\
};
\addplot [color=blue, line width=0.8pt, mark=triangle*, mark options={solid, rotate=180, blue}, mark size=3pt]
  table[row sep=crcr]{%
-10	0.852412151440233\\
-5	0.481110874726176\\
0	0.273662267237067\\
5	0.159302091307707\\
10	0.0985874868732687\\
15	0.0690514630297123\\
20	0.0565892054534286\\
25	0.0520307540951212\\
30	0.0505036903051374\\
};
\addlegendentry{$\text{err}_{\text{pos}}=\unit[0.02]{m},\text{err}_{\text{ori}}=\unit[0]{^\circ}$}

\addplot [color=blue, dashed, line width=1.1pt, forget plot]
  table[row sep=crcr]{%
-10	0.0497816221896763\\
-5	0.0497816221896763\\
0	0.0497816221896763\\
5	0.0497816221896763\\
10	0.0497816221896763\\
15	0.0497816221896763\\
20	0.0497816221896763\\
25	0.0497816221896763\\
30	0.0497816221896763\\
};
\addplot [color=black, line width=0.8pt]
  table[row sep=crcr]{%
-10	0.841193299769191\\
-5	0.473037754932527\\
0	0.266008677973937\\
5	0.149587672483921\\
10	0.0841193299768753\\
15	0.0473037754933275\\
20	0.0266008677973543\\
25	0.0149587672483887\\
30	0.00841193299768136\\
};
\addlegendentry{Mismatch-free}

\end{axis}

 \draw (1.3,2.3) ellipse (0.3 and 0.9 );
\node[right, align=left] at (0.7,1) {\small{Asymptotic PEB}};

\end{tikzpicture}%
    \vspace{-3em}
    \caption{Position error bound (PEB) vs. transmit power with different levels of error in RIS geometry in a \ac{ris}-aided downlink \ac{siso} localization system. This system consists of a single-antenna \ac{bs} at $[5,0,3]^\mathsf{T}$ m, a single-antenna \ac{ue} at $[-2.5,2.5,0]^\mathsf{T} $ m, and a~$64\times 64$ \ac{ris} at $[0,-5,2.5]^\mathsf{T}$ m oriented along the positive $y$-axis. We consider 32 pilot transmissions at a $\unit[28]{GHz}$ carrier frequency, with a $\unit[400]{MHz}$ bandwidth and 3000 subcarriers. The noise PSD is $ \unit[-173.855]{dBm/Hz}$, and the noise figure is $\unit[10]{dB}$. The position error bounds are derived based on the misspecified Cramér-Rao bound (MCRB) theory. Position errors in the RIS are introduced as $\tilde{\pv}=\bar{\pv}+\uv$, where $\uv=\text{err}_\text{pos}\times[1,1,1]^\mathsf{T}$. Orientation errors are introduced as $\tilde{\Rm}=\Rm_v\bar{\Rm}$, with $\Rm_v=\Rm_z(\text{err}_\text{ori})\Rm_y(\text{err}_\text{ori})\Rm_x(\text{err}_\text{ori})$, where $\Rm_z$, $\Rm_y$, and $\Rm_x$ denote the rotation matrices around the Z-axis, Y-axis, and X-axis, respectively.}
    \label{fig2}
\end{figure}
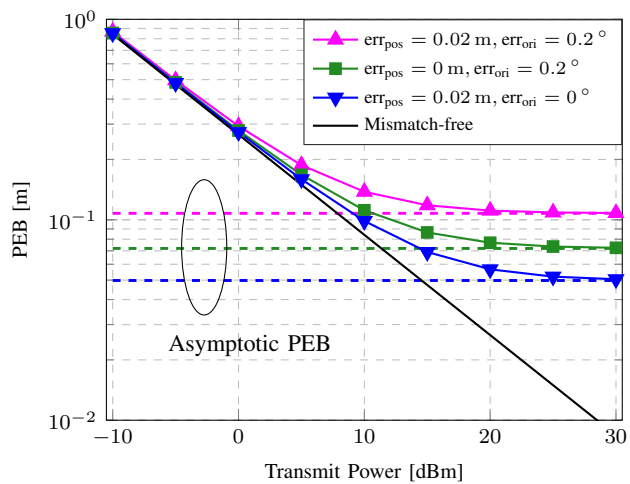

Fig.~\ref{fig2} shows the (M)CRB on the position (the so-called position error bound (PEB)) over different levels of RIS geometry errors. In the mismatch-free case, the localization error depends solely on the SNR of the received signals. A higher SNR (greater transmit power with fixed thermal noise at the receiver) consistently results in a lower PEB. However, in the presence of RIS geometry errors, thermal noise and model mismatch both contribute to localization error. As shown in Fig.~\ref{fig2}, in low-SNR regions, the position error bound aligns with the mismatch-free case, indicating that localization performance is primarily influenced by noise and the impact of model mismatch is negligible. Conversely, in high-SNR regions, localization performance becomes mismatch-dominated and eventually saturates at a certain performance limit (asymptotic PEB). The greater the RIS geometry errors, the more pronounced this effect becomes. This demonstrates that inadequate calibration of RIS geometry can be a significant bottleneck to achieving high-accuracy localization. Additionally, we observe that the localization error can be significantly higher than the calibration error. For example, a $\unit[0.02]{m}$ calibration error in \ac{ris} position can result in a localization error of approximately $\unit[0.05]{m}$ (blue curve). Finally, small RIS orientation errors also lead to large positioning errors when users are far away from the RIS.

\subsubsection{Localization Performance under Hardware Impairments}
In this section, we consider two examples of HWI, including pixel failure and mutual coupling, on the localization performance. Fig.~\ref{fig:LocRMSEvsSNR_Theo} shows the performance degradation in localization due to \textit{pixel failures}. Pixel failures are modeled according to \cite[Eq. 8]{ozturk2024ris} and each element fails independently from each other with probability $p_{\rm{fail}}$. The number of transmission is taken as $20$. Similar to the trends in Fig.~\ref{fig2}, in the low SNR regime, the PEB with failures (represented by LB \cite{ozturk2024ris}) aligns with the accuracy obtained in the absence of failures especially when $p_{\rm{fail}}$ is small. On the other hand, the PEBs with failure reach a plateau in the high SNR regime, suggesting that the impact of failures on localization becomes more pronounced as the SNR increases (leading to almost two orders of magnitude loss compared to the failure-free case when $p_{\rm{fail}} = 2\%$). Hence, pixel failures can be a major showstopper at high SNRs even when only a small number of RIS pixels are faulty. Independent of pixel failures, \textit{strong mutual coupling} can distort the radiation pattern of RISs and increase power consumption. In~\cite{Zheng2024On},  the impact of mutual coupling on channel estimation (which is a key step for both localization and communication) was evaluated (see later in Fig.~\ref{fig3}). It was shown that in uncalibrated cases, closer integration of RIS elements or the enlargement of RIS size can exacerbate mutual coupling and degrade channel estimation performance. Even with perfect calibration of mutual coupling, excessively tight RIS element spacing can substantially impair channel estimation performance.

\subsection{Impact of Model Mismatches on Communication}
In this section, we will demonstrate that communication is generally less affected by model mismatches than localization or sensing. Since this is obvious for geometry mismatches, our focus will be on hardware impairments. We have selected  pixel failure and mutual coupling, for which we present a quantitative analysis. 

\subsubsection{Trade-off Between Communication and Localization under Pixel Failures}
To assess the impact of pixel failures on localization and communication performance, we consider a transmission scheme where out of $T$ total transmissions, $T_p$ are allocated for pilot transmission and the remaining $T_d$ for data transmission. The average transmission power limitations for pilot and data transmissions are denoted as $P_p$ and $P_d$, respectively. These power limits must satisfy the total energy constraint given by $E_{\rm{tot}}$. Moreover, we consider a near-field scenario under LOS blockage, as illustrated through RIS-2 in Fig.~\ref{fig-1-illustration}, and assume the existence of prior knowledge regarding the location of the UE, represented by an uncertainty region. In the pilot phase, the RIS profiles are chosen as positional beams pointing towards positions drawn uniformly from the uncertainty region. The beam that results in the highest power at the UE is chosen in the pilot phase and used throughout the entire data transmission phase, assuming perfect channel estimation. In this setup, localization performance is determined by the $T_p$ pilots while communication performance relies on the $T_d$ transmissions during the data phase.

\begin{figure}[t]
        \subfigure[]{
        \centering
		\begin{tikzpicture}
			\begin{semilogyaxis}[
		      width=2.7in,
            height=2.1in, scale only axis,
            at={(0,0)},
				legend style={nodes={scale= 0.8, transform shape},at={(0,0)},anchor=south west}, 
				legend cell align={left},
                xticklabel style = {font=\color{white!15!black},font=\footnotesize},
                xlabel style={font=\color{white!15!black},font=\footnotesize},
				ylabel={Localization Accuracy [m]},
				xlabel={SNR [dB]},
				xmin=-21, xmax=21,
				ymin=0.01, ymax=10,
				xtick={-20, -15, -10, -5, 0, 5, 10, 15, 20},
				ytick={0.01,0.1,1,10},
                yticklabel style = {font=\color{white!15!black},font=\footnotesize},
                ylabel style={font=\color{white!15!black},font=\footnotesize},
				ymajorgrids=true,
				xmajorgrids=true,
				grid style=dashed,
			    every axis plot/.append style={thick},
				]

            `   \addplot[
				color=magenta,
				mark = square,
				mark options={solid},
                line width=0.8pt,
                mark size = 2.7pt,
				]
				coordinates {
					(-20,4.7506)(-15,2.7932)(-10,1.7691)(-5,1.2859)(0,1.0892)(5,1.0192)(10,0.99601)(15,0.98802)(20,0.98616)
				};

                \addplot[
				color=green,
				mark = o,
				mark options={solid},
                 line width=0.8pt,             
                 mark size = 3pt,
				]
				coordinates {
				    (-20,2.6534)(-15,1.5027)(-10,0.86361)(-5,0.51726)(0,0.34108)(5,0.26173)(10,0.23104)(15,0.22042)(20,0.21698)
				};

				\addplot[
				color=red,
				mark = x,
				mark options={solid},                
                line width=0.8pt,
                mark size = 3pt,
				]
				coordinates {
				    (-20,2.5663)(-15,1.4465)(-10,0.81941)(-5,0.47131)(0,0.28292)(5,0.18738)(10,0.14457)(15,0.12809)(20,0.12242)
				};

            \addplot[
				color=blue,
                style = dashed,
				mark options={solid},
                line width=0.8pt,
				]
				coordinates {
					    (-20,2.7892)(-15,1.5685)(-10,0.88203)(-5,0.496)(0,0.27892) (5,0.15685)(10,0.088203)(15,0.0496)(20,0.027892)
				};
		\legend{$p_{\text{fail}} = 2\%$ (LB), $p_{\text{fail}} = 1\%$ (LB), $p_{\text{fail}} = 0.25\%$ (LB),  Failure-free (PEB)}	
			\end{semilogyaxis}
		\end{tikzpicture}
    \label{fig:LocRMSEvsSNR_Theo}
  }
        \subfigure[]{
        \centering
	\begin{tikzpicture}
			\begin{semilogyaxis}[
		    width=2.7in,
            height=2.1in,scale only axis,
				legend style={nodes={scale= 0.8, transform shape},at={(0,0)},anchor=south west}, 
				legend cell align={left},
				ylabel={Localization Accuracy [m]},
                xlabel={Communication Rate [bps/Hz]},
                xticklabel style = {font=\color{white!15!black},font=\footnotesize},
                xlabel style={font=\color{white!15!black},font=\footnotesize},
                yticklabel style = {font=\color{white!15!black},font=\footnotesize},
                ylabel style={font=\color{white!15!black},font=\footnotesize},
				xmin= 20.1, xmax=20.92,
				ymin=0.01, ymax= 2,
				xtick={20.1, 20.2, 20.3, 20.4, 20.5, 20.6, 20.7, 20.8, 20.9},
				ytick={0.01,0.1,1},
				ymajorgrids=true,
				xmajorgrids=true,
				grid style=dashed,
			    every axis plot/.append style={thick},
				]

            `   \addplot[
				color= magenta,
				mark = o,
				mark options={solid},
                line width=0.8pt,
                mark size = 2.8pt,
                mark indices = {1, 5, 8, 9},
				]
				coordinates {
				(20.822,1.1596)(20.8216,0.96792)(20.8206,0.88417)(20.8178,0.85015)(20.8107,0.83684)(20.7917,0.83174)(20.7406,0.82979)(20.5968,0.82904)(20.1317,0.82875)
				};

                 \addplot[
				color= green,
				mark = x,
				mark options={solid},
                line width=0.8pt,
                mark size = 3pt,
                mark indices = {1, 5, 8, 9},
				]
				coordinates {
				    (20.8169,0.99653)(20.8165,0.83171)(20.8155,0.75969)(20.8128,0.73044)(20.8056,0.71899)(20.7866,0.7146)(20.7355,0.71292)(20.5917,0.71229)(20.1266,0.71203)
				};

            \addplot[
				color= red,
				mark = diamond,
				mark options={solid},
                line width=0.8pt,
                mark size = 3.3pt,
                mark indices = {1, 5, 8, 9},
				]
				coordinates {
				      (20.8488,1.0505)(20.8484,0.68273)(20.8474,0.47293)(20.8447,0.36254)(20.8375,0.31041)(20.8185,0.28814)(20.7674,0.27918)(20.6236,0.27569)(20.1585,0.27435)

				};

            \addplot[
				color= blue,
				mark = +,
				mark options={solid},
                line width=0.8pt,
                mark size = 3pt,
                mark indices = {1, 5, 8, 9},
				]
				coordinates {
                    (20.894,1.0106)(20.8936,0.62327)(20.8925,0.3844)(20.8898,0.23707)(20.8827,0.14621)(20.8637,0.090176)(20.8125,0.055615)(20.6687,0.0343)(20.2037,0.021154)
				};
    
        \legend{$p_{\text{fail}} = 3\%$, $p_{\text{fail}} = 2\%$, $p_{\text{fail}} = 1\%$, $p_{\text{fail}} = 0\%$}	
			\end{semilogyaxis}
		\end{tikzpicture}
         \label{fig:ratevspfail}
		}
	    \caption{Theoretical limits on localization RMSE vs. SNR in \subref{fig:LocRMSEvsSNR_Theo} and theoretical limits on localization RMSE vs. communication rate in \subref{fig:ratevspfail} for various $p_{\rm{fail}}$ values under pixel failures, using the setup from \cite{ozturk2024ris}. 
     An $20\times 20$ RIS is considered. The base station is located at $15\times[-1\, 1\, 1]^\mathsf{T}/\left\lVert[-1\, 1\, 1]\right\rVert$ in \subref{fig:LocRMSEvsSNR_Theo} and at  $10\times[-1\, 1\, 1]^\mathsf{T}/\left\lVert[-1\, 1\, 1]\right\rVert$ in \subref{fig:ratevspfail}. The user is located at $3\times[1\, 1\, 1]^\mathsf{T}/\left\lVert[1\, 1\, 1]\right\rVert$ in \subref{fig:LocRMSEvsSNR_Theo} and at $4\times[1\, 1\, 1]^\mathsf{T}/\left\lVert[1\, 1\, 1]\right\rVert$ in \subref{fig:ratevspfail}.}              \vspace{-0.1in}
        \label{fig:pixel_failure}
\end{figure}
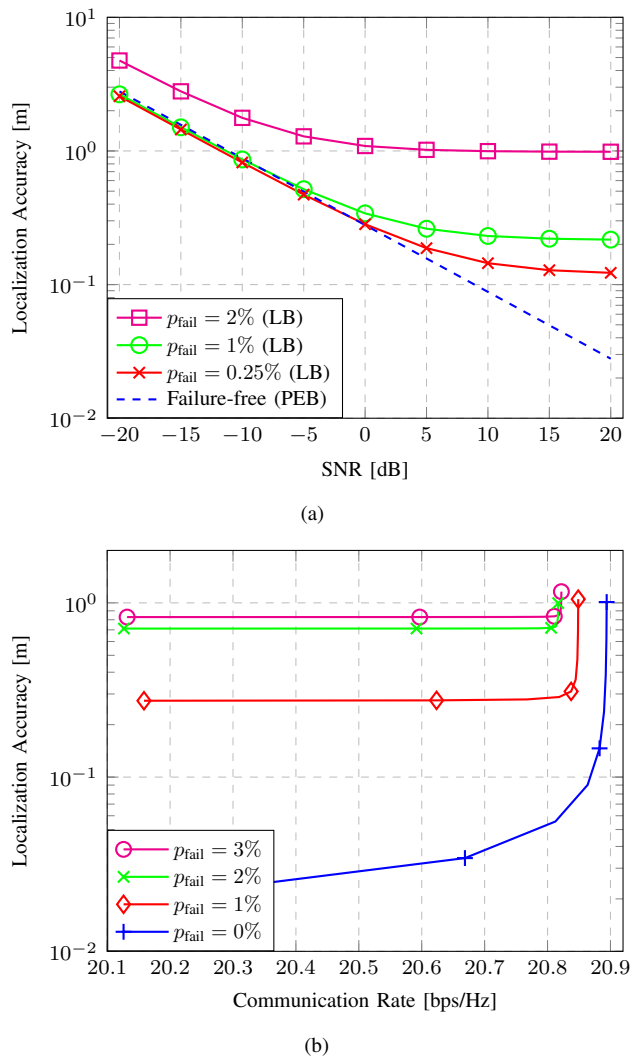

Fig.~\ref{fig:ratevspfail} illustrates the trade-off between localization accuracy (characterized by the LB) and communication rate under pixel failures for several $p_{\rm{fail}}$ values, obtained by sweeping the values of $P_p$ to tune the trade-offs. Here, $T$ is taken as $120$ and $E_{\rm{tot}}$ is set to $100 T$. For each $p_{\rm{fail}}$,  $[-10, 20]\,$ dBW are taken as the possible values of $P_p$. As observed from Fig.~\ref{fig:ratevspfail}, the communication rate remains largely unaffected by pixel failures, consistently achieving values within a very narrow range regardless of $p_{\rm{fail}}$. On the other hand, localization accuracy varies significantly with $p_{\rm{fail}}$, highlighting the large difference in sensitivity to pixel failures between localization and communication. This difference stems from the fact that communication cares about end-to-end composite channel without delving into its geometric structure while localization involves extracting geometric (i.e., location) information from location-dependent phase shifts across the RIS elements over distinct RIS beams during the pilot phase. Hence, unless pixel failures lead to substantial variations in SNR, communication performance remains stable in the face of failures while localization accuracy suffers severely from failure-distorted phase profiles.

\subsubsection{Impact of Inter-element Distance on Communication and Localization}
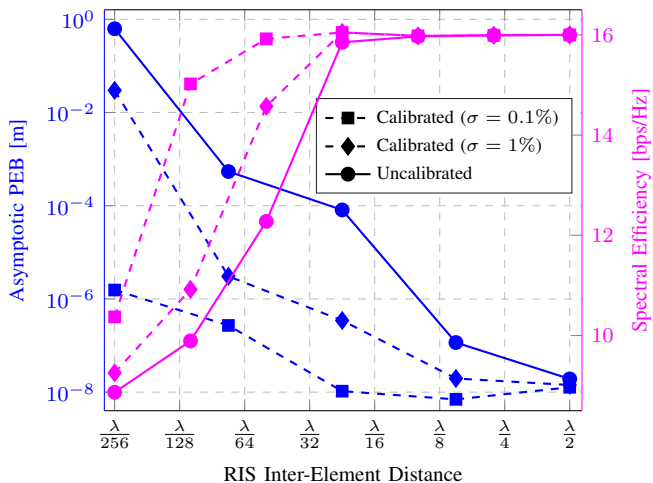
\begin{figure}[t]
    \centering
%
%
\definecolor{Magenta}{rgb}{1.00000,0.00000,1.00000}%
\begin{tikzpicture}

\begin{axis}[%
width=2.5in,
height=2.1in,
at={(0in,0in)},
scale only axis,
xmode=log,
xmin=0.0035,
xmax=0.57,
xminorticks=true,
xticklabel style = {font=\color{white!15!black},font=\footnotesize},
xlabel style={font=\color{white!15!black},font=\footnotesize},
xtick={0.00390625,0.0078125,0.015625,0.03125,0.0625,0.125,0.25,0.5,1},
xticklabels={$\frac{\lambda}{256}$,$\frac{\lambda}{128}$,$\frac{\lambda}{64}$,$\frac{\lambda}{32}$,$\frac{\lambda}{16}$,$\frac{\lambda}{8}$,$\frac{\lambda}{4}$,$\frac{\lambda}{2}$,${\lambda}$},
xlabel={RIS Inter-Element Distance},
ymode=log,
ymin=4e-9,
ymax=1.6,
yticklabel style = {font=\color{white!15!black},font=\footnotesize},
ylabel style={font=\color{white!15!black},font=\footnotesize,color=blue},
axis y line*=left,
y axis line style={blue},
y tick label style={blue},
ylabel={Asymptotic PEB [m]},
axis background/.style={fill=white},
yticklabel pos=right,
xmajorgrids,
xminorgrids,
ymajorgrids,
yminorgrids,
grid style={dashed},
legend style={at={(0.98,0.78)}, anchor=north east, legend cell align=left, align=left, draw=white!15!black, font=\scriptsize}
]
\addplot [color=black, line width=0.8pt, dashed, mark=square*, mark options={solid, black}]
  table[row sep=crcr]{%
100	0.00688311615112336\\
100	2.62092228340806e-07\\
};\addlegendentry{Calibrated ($\sigma=0.1\%$)}
\addplot [color=black, line width=0.8pt, dashed, mark=diamond*, mark options={solid, black}, mark size=3pt]
  table[row sep=crcr]{%
100	0.15335824772616\\
100	0.0728389868273271\\
};\addlegendentry{Calibrated ($\sigma=1\%$)}
\addplot [color=black, line width=0.8pt, mark=*, mark options={solid, black}, mark size=2.4pt]
  table[row sep=crcr]{%
100	0.0335492424210769\\
100	0.00993285102842783\\
};\addlegendentry{Uncalibrated}

\addplot [color=blue, line width=0.8pt, dashed, mark=square*, mark options={solid, fill=blue, draw=blue}, forget plot]
  table[row sep=crcr]{%
0.00390625	1.55451694254192e-06\\
0.0131390064883393	2.70127502138676e-07\\
0.0441941738241592	1.0503269218121e-08\\
0.14865088937534	7.00193643725857e-09\\
0.5	1.2736514472437e-08\\
};
\addplot [color=blue, line width=0.8pt, dashed, mark=diamond*, mark options={solid, fill=blue, draw=blue}, mark size=3pt, forget plot]
  table[row sep=crcr]{%
0.00390625	0.0298652421799625\\
0.0131390064883393	3.06032240834848e-06\\
0.0441941738241592	3.45071718741255e-07\\
0.14865088937534	1.96757732761285e-08\\
0.5	1.42394629914529e-08\\
};
\addplot [color=blue, line width=0.8pt, mark=*, mark options={solid, fill=blue, draw=blue}, mark size=2.4pt, forget plot]
  table[row sep=crcr]{%
0.00390625	0.627486117102375\\
0.0131390064883393	0.000542137697365278\\
0.0441941738241592	8.12491680408469e-05\\
0.14865088937534	1.16102949361429e-07\\
0.5	1.92543742577959e-08\\
};
\end{axis}

\begin{axis}[%
width=2.5in,
height=2.1in,
at={(0in,0in)},
scale only axis,
xmode=log,
xmin=0.0035,
xmax=0.57,
axis x line=none,
separate axis lines,
ymin=8.5,
ymax=16.5,
yticklabel style = {font=\color{white!15!black},font=\footnotesize},
ylabel style={font=\color{white!15!black},font=\footnotesize,color=Magenta},
axis y line*=right,
y axis line style={Magenta},
y tick label style={Magenta},
ylabel={Spectral Efficiency [bps/Hz]},
yticklabel pos=right
]
\addplot [color=Magenta, line width=0.8pt, dashed, mark=square*, mark options={solid, Magenta}, forget plot]
  table[row sep=crcr]{%
0.00390625	10.3722236117873\\
0.00876923475241698	15.0217806416103\\
0.0196862664046074	15.9212510750716\\
0.0441941738241592	16.048362409928\\
0.0992125657480125	15.9809021246667\\
0.222724679535085	15.9917820143265\\
0.5	15.9993219994624\\
};
\addplot [color=Magenta, line width=0.8pt, dashed, mark=diamond*, mark size=3pt, mark options={solid, Magenta}, forget plot]
  table[row sep=crcr]{%
0.00390625	9.25440123109995\\
0.00876923475241698	10.9209188654909\\
0.0196862664046074	14.5731174955849\\
0.0441941738241592	16.0489176990069\\
0.0992125657480125	15.9809093331055\\
0.222724679535085	15.991783567763\\
0.5	15.9993161686131\\
};
\addplot [color=Magenta, line width=0.8pt, mark=*, mark options={solid, Magenta}, mark size=2.4pt, forget plot]
  table[row sep=crcr]{%
0.00390625	8.86709067284951\\
0.00876923475241698	9.89251272595571\\
0.0196862664046074	12.2784795415671\\
0.0441941738241592	15.85165123657\\
0.0992125657480125	15.9701745471621\\
0.222724679535085	15.9917099054654\\
0.5	15.9992905636414\\
};
\end{axis}
\end{tikzpicture}%
    \vspace{-3em}
    \caption{Asymptotic PEB (localization performance, as defined in Fig.~\ref{fig2}) and spectral efficiency (communication performance) vs. inter-element distance of the RIS in a RIS-aided SIMO system (consisting of a single-antenna transmitter, a 16-antenna receiver, and a 64-element passive RIS) operating at 30 GHz. Both the RIS and the antenna array are configured as uniform linear arrays. We adjust the inter-element distance of the RIS while fixing the BS's antenna array with a half-wavelength spacing. The coupling parameters are obtained by first computing Z-parameters according to~\cite[Eq. (2)]{Zheng2024On} and then transforming them to S-parameters. Here, $\lambda$ denotes the signal wavelength. A shorter inter-element distance typically results in stronger mutual coupling. The calibrated cases assume noisy prior knowledge of each S-parameter, with the relative residual error represented by~$\sigma$. For localization, random beamformers are employed, whereas for communication, optimal beamforming is performed based on~$\hat{S}$ and perfect channel state information using a successive convex approximation-based optimization algorithm.} 
    \label{fig3}
\end{figure}

Fig.~\ref{fig3} illustrates the performance of a joint localization and communication system aided by a passive RIS with mutual coupling. In this RIS-aided SIMO communication setup, the asymptotic PEB (localization metric, the horizonal lines in Fig.~\ref{fig2}) and spectral efficiency (communication metric) are evaluated across various inter-element distances of the RIS. Fig.~\ref{fig3} reveals that a reduction in the inter-element distance of the RIS causes stronger mutual coupling, which in turn degrades both localization and communication performance. However, after precise calibration, these performance losses due to mutual coupling can be effectively mitigated. The more accurate the calibration, the better the performance of both localization and communication. Additionally, it is observed that localization is more sensitive to RIS mutual coupling compared to communication. However, this also implies that the influence of mutual coupling on communication could be even more significant than our assessment assuming perfect channel state information, since inaccurate localization typically indicates challenges in achieving precise channel estimation and beamforming, thereby further degrading communication performance. These observations emphasize the necessity of accurate calibration.

\subsection{Effect of RIS Architectures}\label{Effect_RIS}
In this section, we discuss the effect of RIS architectures on the RIS-aided ISAC system in terms of calibration. All RIS types are affected by geometry mismatches. For instance, when angle-of-arrival (AOA) and angle-of-departure (AOD) measurements are used, orientation mismatches play an important role. In contrast, with time-of-arrival (TOA)-only measurements, orientation mismatches primarily affect the SNR. Additionally, hardware mismatches can vary among different RIS types due to variations in hardware. However, pixel failures and reflection coefficient (or phase shifter) impairments are a more common hardware mismatch across different RIS types.  Pixel failures mainly affect SNR for non-coherent methods, while they lead to severe degradations in coherent methods. 

For \textit{passive RIS}, TOA and AOA/AOD information at the user or network-side can be used, while in \textit{hybrid RIS} (HRIS), 
geometry calibration is significantly simplified, thanks to the 
sensing capability of the HRIS~\cite{ghazalian2023joint}. Empowered by reflection-type amplifiers, \textit{active RISs} can simultaneously shift signal phases and boost signal amplitudes, enhancing both localization and communication through improved signal reception. However, active RISs also introduce additional thermal noise and exacerbate mutual coupling, potentially amplifying model mismatch and compromising localization and communication performance. As revealed in~\cite{zheng2023jrcup}, an optimal amplification factor exists when considering this double-edged sword effect of active RISs. In addition, the introduction of the active amplifier can further increase the phase-dependent amplitude variation. Recently, \textit{beyond diagonal-RIS} (BD-RIS) has been proposed by modeling a RIS as multiple antennas connected to a multiport reconfigurable impedance network, providing performance improvements in various scenarios due to its enhanced design flexibility. The concept of BD-RIS can be employed for both reflective and STAR-type  RISs, as well as for multi-sector RISs. However, calibrating a BD-RIS is more challenging than the conventional RIS due to the increased number of parameters resulting from inter-port connections. In particular, BD-RIS can lead to more hardware failures as connections between RIS elements may be faulty. Finally, STAR-RIS also has its own hardware impairments, such as mutual coupling between the two layers (one for transmission and one for reflection) in a two-layer design. In the case of a single-layer STAR-RIS, the phase adjustment of an element is more prone to impairment compared to other RIS types, requiring continuous calibration, which impacts both transmission and reflection. Therefore, due to these hardware impairments, calibrating the STAR-RIS is more challenging than passive, active, and hybrid RISs. 

\vspace{3mm}

\section{Calibration Methodologies}
In this section, we describe two main calibration types that should be used in the \ac{isac} system. Additionally, we briefly outline BS calibration in the positioning system based on the 3rd Generation Partnership Project (3GPP) specification. We also describe joint RIS calibration and user localization systems. Finally, we summarize two calibration methodologies: model-based (described in described in sections III-A and III-B) and learning-based (given in Sec.~\ref{Learning_basedmodel}) in Table~\ref{table}. As shown, we outline the pros and cons of each approach, the requirements for their implementation, and the potential scenarios where each methodology can be applied. 
\subsection{Geometry Calibration} \label{Geo-Cal-Method}
\subsubsection{Standard Approaches}
For BS calibration in uplink (UL) positioning systems, a reference user equipment (UE) with a known location transmits a Sounding Reference Signal (SRS) to the base station (BS). The BS then estimates its own location and compares it with the true location to determine correction terms for other UEs~\cite[Ch.5.4.5]{3gpp38305}. For downlink (DL) positioning calibration, the reference UE performs measurements (e.g., TOA and AOA) and reports these to the BS. A similar process is used for UL positioning in the DL. 
\subsubsection{RIS-aided Systems}
In a RIS-aided \ac{isac} system, RIS geometry calibration is required in addition to BS calibration.  During this process, it is assumed that the hardware impairment parameters are known. Note that RIS geometry calibration can be an on-demand or continuous process. For movable RIS, this process should be done continuously. However, when the RIS is deployed for the first time or re-installed after maintenance, calibration can be performed on demand. RIS geometry calibration can be carried out using a multi-static or bi-static sensing system in either UL or DL scenarios in TDD mode. In this system, certain UE anchors (reference UEs) with known states transmit or receive signals to/from the BS via a RIS. Based on the received signals at the BS or reference UEs, measurements can be extracted according to their capabilities. For instance, anchors capable of transmitting or receiving high-bandwidth signals can measure TOA. Additionally, anchors equipped with arrays can capture AOA/AOD measurements. These measurements can be estimated using the maximum-likelihood estimator and other techniques such as compressed sensing, MUSIC, and ESPRIT. The RIS state is then estimated by establishing a triangulation relationship among the BS, reference UEs, and the RIS, based on TOA and AOA/AOD measurements. Geometry calibration can be performed in either a near-field (NF) or far-field (FF) regime, with NF calibration being feasible but requiring more complex algorithms.

The aforementioned approach can \textit{calibirate the RIS geometry and estimate the UEs locations jointly}. The feasibility of such a system depends on having a sufficient number of measurements relative to the number of unknown states, meaning that the number of geometric observations must exceed the number of geometric unknowns. In this system, known as joint RIS geometry calibration and UE localization~\cite{zheng2023jrcup}, the RIS can utilize any of the previously mentioned architectures. However, having sensing capability at the RIS can provide additional measurements, thereby reducing the complexity of the localization algorithm.

\subsection{Hardware Calibration} \label{HWI-Cal-Method}
\subsubsection{Standard Approaches}
For BS calibration in positioning systems, certain hardware calibrations, such as antenna calibration and beam pattern measurements, are performed offline in a factory setting, often within an anechoic chamber. However, online calibration is also necessary due to factors such as temperature changes during normal operation, which can affect the frequency response of the antenna system and impact BS performance. This can be achieved using a reference UE, as in geometry calibration.

\subsubsection{RIS-aided Systems}
RIS hardware calibration can be performed offline, just like BS calibration. For instance, RIS calibration to address hardware impairments such as mutual coupling can be conducted offline, although this process is time-consuming and may even be impractical once the RIS is deployed. Additionally, hardware failures (e.g., pixel failures) can occur during normal operation. Therefore, online hardware calibration is required after system deployment and throughout normal operation for regular hardware monitoring. Generally, online hardware calibration can be approached by estimating unknown parameters based on a signal model, where the geometric state of the anchor is assumed to be known. This calibration system can be categorized into agent-based and self-calibration methods. Agent-based calibration requires additional effort to determine the hardware state of the agent but simplifies the calibration algorithm. In contrast, self-calibration offers convenience but may result in reduced calibration performance. Agent-based RIS hardware calibration follows a similar approach to the geometry RIS calibration described in Sec.~\ref{Geo-Cal-Method}, except that the RIS has a known geometric state.

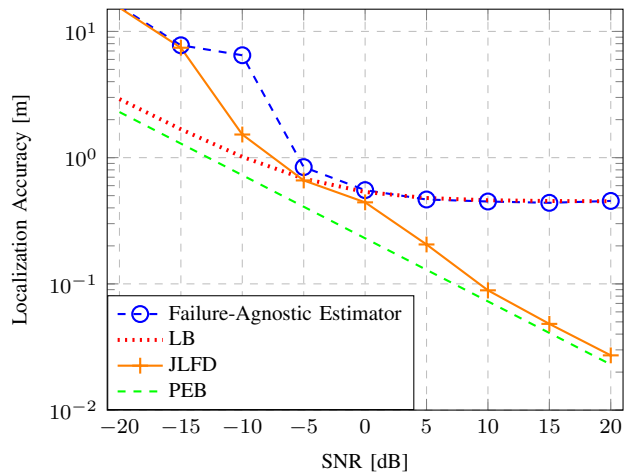
\begin{figure}
		\centering
		\begin{tikzpicture}
			\begin{semilogyaxis}[
		      width=2.7in,
            height=2.1in,scale only axis,
				legend style={nodes={scale= 0.8, transform shape},at={(0,0)},anchor=south west}, 
				legend cell align={left},
                xticklabel style = {font=\color{white!15!black},font=\footnotesize},
                xlabel style={font=\color{white!15!black},font=\footnotesize},
                yticklabel style = {font=\color{white!15!black},font=\footnotesize},
                ylabel style={font=\color{white!15!black},font=\footnotesize},
				ylabel={Localization Accuracy [m]},
				xlabel={SNR [dB]},
				xmin=-21, xmax=21,
				ymin=0.01, ymax=15,
				xtick={-20, -15, -10, -5, 0, 5, 10, 15, 20},
				ytick={0.01,0.1,1,10},
				ymajorgrids=true,
				xmajorgrids=true,
				grid style=dashed,
			    every axis plot/.append style={thick},
				]

                 \addplot[
				color= blue,
                line width=0.8pt,
                mark = o,
				mark options={solid},
                mark size = 3pt,
                style = dashed,
				]
				coordinates {
				        (-20,15.8136)(-15,7.7648)(-10,6.4755)(-5,0.84138)(0,0.55292)(5,0.46604)(10,0.44926)(15,0.43868)(20,0.45396)
				};

                \addplot[
                color = red,
                style = dotted,
                line width=1.5pt,
				]
				coordinates {
				       (-20,2.9113)(-15,1.6788)(-10,1.0149)(-5,0.68151)(0,0.53442)(5,0.47873)(10,0.45966)(15,0.45339)(20,0.45142)

				};
                 \addplot[
				color= orange,
		          mark = +,
				mark options={solid},
                line width=0.8pt,
                mark size = 3pt,
				]
				coordinates {
				    (-20,15.4085)(-15,7.4238)(-10,1.525)(-5,0.6601)(0,0.44249)(5,0.20525)(10,0.088897)(15,0.048198)(20,0.027145)
				};

            `   \addplot[
				color= green,
		          style = dashed,
				mark options={solid},
                line width=0.8pt,
				]
				coordinates {
				     (-20,2.2975)(-15,1.292)(-10,0.72654)(-5,0.40857)(0,0.22975)(5,0.1292)(10,0.072654)(15,0.040857)(20,0.022975)
				};

            \legend{Failure-Agnostic Estimator,  LB,  JLFD, PEB}	

			\end{semilogyaxis}
		\end{tikzpicture}
	   \caption{Localization RMSE vs. SNR obtained by the failure-agnostic estimator and the joint localization and failure diagnosis (JLFD) estimator along with the corresponding theoretical bounds when $p_{\rm{fail}} = 1\%$. PEB provides a theoretical limit on localization accuracy when the user has perfect knowledge of pixel failure mask. The same setup as in Fig.~\ref{fig:LocRMSEvsSNR_Theo} is considered.}
    \label{fig:pixel_failure_pfail_1_perf}
	\end{figure}

Agent-based \emph{RIS hardware calibration can be conducted jointly with UE localization}. For instance, in the context of pixel failures, joint localization and failure diagnosis (JLFD) can be performed, where the UE diagnoses RIS pixel failures as part of the hardware calibration while simultaneously estimating its location~\cite{ozturk2024ris}. Fig.~\ref{fig:pixel_failure_pfail_1_perf} illustrates the performance of JLFD along with the theoretical bounds on the localization when $p_{\rm{fail}} = 1\%$. The performance is compared to  failure-agnostic estimator (which ignores the presence of failures) saturates at high SNRs (attaining the corresponding bound quantified by the LB) since the mismatch between the true model with failures and the ideal model without failures becomes the dominating factor. On the other hand, the JLFD algorithm can largely close the performance gap to the ideal case without failures (quantified by the PEB) by effectively taking into account the presence of failures. 

\subsection{Learning-based Calibration}\label{Learning_basedmodel}

\begin{table*}
\centering
\label{table}
\caption{Summary of RIS Calibration Methodologies}
\renewcommand{\arraystretch}{1.7}
\begin{tabular}{m{0.1\textwidth}<{\centering} | m{0.13\textwidth}<{\centering} | m{0.15\textwidth}<{\centering} m{0.15\textwidth}<{\centering} m{0.13\textwidth}<{\centering} m{0.17\textwidth}<{\centering}  }
  \hthickline 
  \multicolumn{2}{c|}{\textbf{Methodology}} & \textbf{Scenarios} & \textbf{Requirements} & \textbf{Pros} & \textbf{Cons} \\
  \hthickline
  \multirow{2}{*}[-1.5ex]{Model-based} & Standard calibration & Work for both geometry and hardware & Known calibration model and agent state & Low complexity & High cost with the agent, accurate model needed\\
  \cline{2-6}
      & Joint
localization and calibration & Work for both geometry and hardware & Known calibration model & Practical onsite calibration & High complexity or extra processing unit needed \\
\hline
\multirow{2}{*}[-1.5ex]{Learning-based} & Fingerprinting & Learn the signal-to-state mapping & Labeled dataset & Calibrates all parameters at once & Large dataset required, less scalable, sensitive to environment change\\
\cline{2-6}
      & Model-driven
DL & Learn the calibration parameters & Labeled dataset and calibration model & Small dataset, fast training & Accurate model needed \\
  \hthickline 
\end{tabular}
\end{table*}

{The aforementioned calibration methods are typically used when accurate signal and hardware models are available in RIS-aided ISAC systems and when these models are simple.
However, when an accurate model is not available, is overly complicated, or when dealing with an accurate generative model with unknown parameters, machine learning methods are preferred~\cite{ma2022learn}.

\subsubsection{Fingerprinting}
When we lack information about the channel model, we can use black-box deep learning. In fingerprinting-based localization, a supervised machine learning-based approach learns a function mapping the observed signal to the UE location. In this case, the calibration process is naturally integrated into the training phase to learn the true model. However, such pure data-driven approaches often require extensive training data to capture the nuances of hardware impairments and environmental conditions, which may not always be available in practical scenarios. Another critical concern is overfitting, where a calibration model tailored too closely to specific training data may perform poorly under slightly different operational conditions.

\subsubsection{Model-driven Deep Learning}
When we have a good generative model, but with unknown parameters, model-based deep learning can be applied. In~\cite{ma2022learn}, the calibration model is integrated into the learning framework~\cite{ma2022learn}. On the one hand, the complexity of calibration models can be reduced through a learning-based surrogate model, which extracts model features and decreases the number of measurement cycles needed for calibration~\cite{zhou2022transfer}. On the other hand, a sufficient number of measurements is required for classical calibration methods. Additionally, high-dimensional calibration parameters can be included as part of the trainable neural network parameters, which can be conveniently optimized within machine learning frameworks. In this context, operation over real channels is crucial. Supervised learning and reinforcement learning can be applied to receiver learning and transmitter learning, respectively. 

The implementation of learning-based calibration in RIS-aided systems does not make a huge difference to BS calibration. However, the online optimization of RIS profiles may introduce extra training complexity. Thus, the appropriate learning framework should be selected, balancing the benefits of improved performance against the costs of implementation and maintenance. 

\section{Challenges and Future Directions}
 
In this paper, we have demonstrated that RIS calibration is crucial for unlocking ISAC capabilities. Both geometric and hardware calibration are required to avoid severe performance penalties for communication, but especially for localization and sensing. Therefore, we argue that all RIS-based ISAC work should take calibration aspects into account. This paper has only addressed a few selected topics in this area, and many challenges remain unsolved.
\begin{enumerate}
    \item \textit{Weak signal of the RIS path:} 
    The calibration is conducted based on measurements obtained via the reflected path from the RIS to the BS, which tends to be weak. This presents a significant challenge in RIS calibration. One potential solution is to amplify the reflected path (i.e., using active RIS). However, such a technique comes with its own challenges (as discussed in Sec.~\ref{Effect_RIS}). Alternatively, utilizing the NF ISAC system for RIS calibration offers a promising solution to address poor signal strength and provides more robust calibration procedures, albeit at the expense of more complex algorithms. Finally, intentionally blocking the direct path can help tackle this issue, although the direct path is useful for clock bias , especially when joint RIS calibration and UE localization are considered.
    
    \item \textit{Extra large RIS:} Online joint localization and calibration becomes challenging with increasing RIS sizes. Detecting impairments in individual RIS elements relying on few transmissions in a coherent processing interval requires sophisticated, high-complexity algorithms and realistic modeling of RIS hardware. Future research should focus on low-complexity and data-driven calibration algorithms to circumvent the issues of RIS model deficiency. 
    \item \textit{Cooperative calibration:} Sidelink communication between UEs can extend communication coverage, which is also helpful for localization and sensing. Additional information between collaborative UEs can reduce the time for calibration. Moreover, cooperative positioning can provide more calibration agents, while multiple positions of the UEs can improve performance (e.g., reduced blind areas and better beam calibration on different angles). Coordination and resource management are the key to realize cooperative calibration. 

    \item \textit{Complex environments:} 
    Calibration can be viewed as an estimation problem integrated with sensing tasks. However, in real-world scenarios, the presence of multipath effects introduces additional unknowns, significantly increasing the complexity of calibration. This means that joint calibration and localization tasks must account for multipath in these scenarios. Yet, with a detailed map of the surrounding environment, these multipath effects can be leveraged to aid in calibration. Specifically, a strong reflector with a known state can be used to counteract the impact of that path and even provide geometric information to enhance positioning accuracy.
\end{enumerate}

\balance 
\bibliographystyle{IEEEtran}
\bibliography{ref}

\vspace{-0.5cm}
\begin{IEEEbiographynophoto}{Reza Ghazalian} (reza.ghazalian@nokia.com) He is a senior RF Specification Engineer at Nokia Mobile Network, Espoo, Finland. His research interests include RIS, radio localization, signal processing and optimization.  
\end{IEEEbiographynophoto}

\vspace{-0.5cm}
\begin{IEEEbiographynophoto}{Pinjun Zheng}
(pinjun.zheng@kaust.edu.sa) is a PhD candidate at KAUST, Saudi Arabia. His research interests include signal processing for 5G/6G localization and communication applications.
\end{IEEEbiographynophoto}

\vspace{-0.5cm}
\begin{IEEEbiographynophoto}{Hui Chen}
(hui.chen@chalmers.se) is a postdoctoral researcher at Chalmers University of Technology, 41296 Gothenburg, Sweden. His research interests include mmWave/THz and RIS-aided localization. \end{IEEEbiographynophoto}

\vspace{-0.5cm}
\begin{IEEEbiographynophoto}{Cuneyd Ozturk}
(cuneyd93@gmail.com) is a lead engineer at Aselsan Inc., Ankara, Turkey. His research interests include satellite communication networks and RIS-aided localization. \end{IEEEbiographynophoto}

\vspace{-0.5cm}
\begin{IEEEbiographynophoto}{Musa Furkan Keskin}
(furkan@chalmers.se) is a research specialist at Chalmers University of Technology, Gothenburg, Sweden. His current research interests include integrated sensing and communications, RIS-aided localization and sensing, and hardware impairments in beyond 5G/6G systems.
\end{IEEEbiographynophoto}

\vspace{-0.5cm}
\begin{IEEEbiographynophoto}{Vincenzo~Sciancalepore}
(vincenzo.sciancalepore@neclab.eu) is a Principal Researcher at NEC Laboratories Europe, Germany, focusing his activity on RISs. He is an editor of IEEE Transactions on Wireless Communications and IEEE Transactions on Communications.
\end{IEEEbiographynophoto}

\vspace{-0.5cm}
\begin{IEEEbiographynophoto}{Sinan~Gezici}
(gezici@ee.bilkent.edu.tr) is a professor at Bilkent University, Ankara, Turkey. His research interests include statistical signal processing, visible light systems, and wireless localization.
\end{IEEEbiographynophoto}

\vspace{-0.5cm}
\begin{IEEEbiographynophoto}{Tareq~Y.~Al-Naffouri}
(tareq.alnaffouri@kaust.edu.sa) is a professor at KAUST, Saudi Arabia. His research interests include sparse, adaptive, and statistical inference/learning and their applications to wireless communications, localization, smart cities, and smart health.
\end{IEEEbiographynophoto}

\vspace{-0.5cm}
\begin{IEEEbiographynophoto}{Henk Wymeersch} (henkw@chalmers.se)
is a professor at Chalmers University
of Technology, 41296 Gothenburg, Sweden. His research interests include 5G
and beyond 5G radio localization and sensing.
\end{IEEEbiographynophoto}

\vfill

\end{document}